\documentclass[showpacs]{revtex4}
\usepackage{amsmath}

\begin{document}

\makeatletter
\def\@cite#1#2{\textsuperscript{[{#1\if@tempswa #2\fi}]}}
\makeatother

\title{Production of squeezed state of single mode cavity field by the
coupling of squeezed vacuum field reservoir in nonautonomous
case\footnote{Supported in part by the National Natural Science
Foundation under grants No.10175029 and 10004012, the Doctoral
Education Fund of the Education Ministry and Post-doctoral Science
Foundation, and the Nuclear Theory Research Fund of HIRFL of
China.}}
\author{Jun-Hong An$^{1}$, Shun-Jin Wang$^{1,2}$\footnote{%
The corresponding author S. J. Wang's Email address:
gontp@lzu.edu.cn}, Hong-Gang Luo$^{1}$ and Cheng-Long Jia$^{1}$}
\affiliation{1. Department of Modern Physics of Lanzhou University,
Lanzhou 730000, P. R.
China\\
2.Department of Physics of Sichuan University, Chengdu 610065, P.
R. China }
\begin{abstract}
The dissipative and decoherence properties as well as the
asymptotic behavior of the single mode electromagnetic field
interacting with the time-dependent squeezed vacuum field
reservoir are investigated in detail by using the algebraic
dynamical method. With the help of the left and right
representations of the relevant $hw(4)$ algebra, the dynamical
symmetry of the nonautonomous master equation of the system is
found to be $su\left( 1,1\right)$. The unique equilibrium steady
solution is found to be the squeezed state and any initial state
of the system is proved to approach the unique squeezed state
asymptotically. Thus the squeezed vacuum field reservoir is found
to play the role of a squeezing mold of the cavity field.
\end{abstract}

\pacs{03.65.Fd, 42.50.Dv, 42.88.+h}

\maketitle

The fundamental property of the squeezed state is that the quantum
fluctuations in one quadrature component of the field can be reduced
heavily. Following from the early work \cite{pleb,yue} on the
squeezed state, researchers have taken great effort on this specific
state. The first experimental result for the generation of the
squeezed state was reported by Slusher $et$ $al$.\cite{slu} with the
scheme of 4-wave mixing in an optical cavity. Currently the
successful scheme for generating squeezed light can also be based on
the parametric oscillator or parametric down converter \cite{pozik}.
Recently, due to its potential applications in the fields of quantum
measurement, optical communication, and quantum information,
squeezed vacuum state has been extensively studied \cite{Fur,Hau}. A
natural problem is what effect of physical systems can be induced by
the squeezed vacuum. The squeezed light field will generally be
characterized as a non-stationary reservoir which contains phase
dependent features, i.e. non-zero values for correlation function
between pairs of photons. When the bandwidths of the squeezed light
are not too small, it can be treated as a Markovian reservoir, and
the master equations of the reduced systems can be obtained based on
Markovian approximation \cite{Scu}.

Master equations, which are derived by integrating out the enormous
irrelevant degrees of freedom of the environment, are of fundamental
importance on the treatment of open quantum systems. The common
feature of the quantum master equations is the existence of the
sandwich terms of the Liouville operators where the reduced density
matrix of the system is between some quantum excitation and
de-excitation operators. Thus, it is very difficult to get the exact
analytical solution of the master equation, only some simple cases
such as a single mode of cavity field coupled a vacuum state
reservoir ($T=0$) or stationary regime properties are normally
treated \cite{Scu}, where the master equations are normally
converted into c-number equations in the coherent state
representation--the Fokker-Planck equation \cite{Gar,Wall}.

In our previous work \cite{Zhao,Wang01,An,An2}, we have proposed and
developed an algebraic method to treat the sandwich terms in the
Liouville operator for the nonequilibrium quantum statistical
systems. This method is just a generalization of the algebraic
dynamical method \cite{Wang93} from quantum mechanical systems to
quantum statistical systems. According to the characteristics of the
sandwich terms in the Liouville operator, the left and right
representations \cite{Wang89} have been introduced and the
corresponding composite algebra has been constructed. As a result,
the master equation is converted into a Schr\"{o}dinger-like
equation and the problems can be solved exactly. Using this method,
we have successfully solved the von Nuemann equation of the quantum
statistical characteristic function of the two-level Janes-Cummings
model \cite{Zhao}, the master equations of sympathetic cooling of
Bose-Einstein condensate in the mean field approximation
\cite{Wang01}, the nonautonomous dissipative two-level atom system
\cite{An}, and the nonautonomous dissipative two identical cavity
system \cite{An2}. The salient feature of our method resides in the
treatment of nonautonomous systems which are related to the control
of man-made quantum systems by changing the coupling constant or
adjusting other parameters of the system. It would be very desirable
to get the analytical solution of the corresponding master equation
of such a system. The nonautonomous system is the one whose
Hamiltonian is time-dependent through some parameters employed for
the purpose of control of the system. However, even if the total
Hamiltonian of a composite system-the system investigated plus the
environment, is time-independent, the master equation of the reduced
density matrix of the investigated system still becomes
nonautonomous under the non-Markovian dynamics \cite{Ana}.
Therefore, the quantum master equation of the reduced density
matrix, in general, should be nonautonomous.

In this letter, using the algebraic dynamical method, we shall solve
the problem of single mode electromagnetic field interacting with
the squeezed vacuum field reservoir and investigate what effect on
the system can be induced by the squeezed vacuum field reservoir. As
is well known, as the environment is ordinary thermal equilibrium
fields, the systems have been investigated very well. When the
environment is the squeezed vacuum field, some work has been done
for the autonomous case \cite{Gea,Dung,Vid}. For the autonomous
case, the investigation of the single mode electromagnetic field in
the squeezed vacuum field can be found in \cite{Dung} where the
master equation was converted into a Fokker-Planck equation, and in
\cite{Vid} where the general solution of the master equation was
given using a superoperator technique. However, for the system being
nonautonomous, to our knowledge, no investigation exists up to now.
It is our goal to study such a nonautonomous case. By using the
algebraic dynamical method, here the $su(1,1)$ dynamical symmetry of
the nonautonomous master equation for the single mode
electromagnetic field is explored and the system is thus proved to
be integrable for the first time. The corresponding analytical
solutions, both the steady solutions and the time-dependent
solutions of the systems are obtained exactly. Any time-dependent
solution of the systems is proven to approach the unique steady
equilibrium solution and the squeezed vacuum is found to play the
role of the squeezing mold of the single mode cavity field.

\bigskip The Hamiltonian of the open system reads as%
\begin{equation*}
H=\hbar \omega (a^{\dagger }a+\frac{1}{2})+\hbar \sum_{k}\omega
_{k}(a_{k}^{\dagger }a_{k}+\frac{1}{2})+\hbar \sum_{k}g_{k}(a^{\dag
}a_{k}+h.c.),
\end{equation*}%
which describes the single mode cavity field interacting with the
squeezed vacuum field reservoir. The total initial density operator
is \cite{Scu}
\begin{equation*}
\rho _{T}=\rho \left( 0\right) \otimes \prod_{k}S_{k}\left( \xi \right)
|0_{k}\rangle \langle 0_{k}|S_{k}^{\dagger }\left( \xi \right) ,
\end{equation*}%
where $S_{k}\left( \xi \right) $ is the squeezing operator and reads
as
\begin{equation*}
S_{k}\left( \xi \right) =\exp (\xi ^{\ast }a_{k_{0}+k}a_{k_{0}-k}-\xi
a_{k_{0}+k}^{\dagger }a_{k_{0}-k}^{\dagger })
\end{equation*}%
with $\omega =ck_{0}$ and $\xi =r\exp \left( i\theta \right) $, $r$ being
the squeezing parameter and $\theta $ being the reference phase for the
squeezed field. The master equation of the reduced density operator for the
field can be obtained in the usual way%
\begin{eqnarray}
\dot{\rho} &=&\frac{\gamma (t)}{2}\left( N+1\right) (2a\rho a^{\dagger
}-a^{\dagger }a\rho -\rho a^{\dagger }a)  \notag \\
&&+\frac{\gamma (t)}{2}N(2a^{\dagger }\rho a-aa^{\dagger }\rho -\rho
aa^{\dagger })  \notag \\
&&-\frac{\gamma (t)}{2}M(2a\rho a-aa\rho -\rho aa)  \notag \\
&&-\frac{\gamma (t)}{2}M^{\ast }(2a^{\dagger }\rho a^{\dagger }-a^{\dagger
}a^{\dagger }\rho -\rho a^{\dagger }a^{\dagger }),  \label{master}
\end{eqnarray}%
where $\langle a_{k}^{\dagger }a_{k^{\prime }}\rangle =N\delta _{kk^{\prime
}}=\sinh ^{2}(r)\delta _{kk^{\prime }}$ and $\langle a_{k}^{\dagger
}a_{k^{\prime }}^{\dagger }\rangle =-M\delta _{k^{\prime },2k_{0}-k}=-\cosh
\left( r\right) \sinh \left( r\right) \exp \left( -i\theta \right) \delta
_{k^{\prime },2k_{0}-k}$, corresponding to $N$ is very large. When $%
N\rightarrow \bar{n}=\frac{1}{\exp (\frac{\hbar \nu _{k}}{k_{B}T})-1}$, $%
M\rightarrow 0$, the master equation(\ref{master}) reduces to the
familiar form of the master equation which describes the system
coupling to the ordinary thermal equilibrium radiation field. The
last two terms of the above equation exhibit the phase-sensitive
nature of the investigated system. In the model considered above, we
assume that one can control the squeezed field so that the decay
rate $\gamma $ can be changed and is time-dependent. The system thus
becomes nonautonomous. Eq. (\ref{master}) in the autonomous case has
been applied to the reduction of laser noise through injection of
the squeezed vacuum \cite{Gea} and to the noise-free phase-sensitive
amplification via the two-photon correlated-emission-laser
\cite{Scu2}.

\bigskip

To explore the algebraic structure and dynamical symmetry of Eq.
(\ref{master}) in the nonautonomous case, we make a squeeze
transformation or a general canonical Bogoliubov transformation
\cite{Bogo} and transform the electromagnetic field operators
$(a,a^{+})$ to the squeezed ones $(A,A^{+})$ as follows
\begin{eqnarray}
A &=&\cosh \left( r\right) a-\sinh \left( r\right) \exp \left( i\theta
\right) a^{\dagger }  \notag \\
&=&S^{\dagger }aS,  \notag \\
A^{\dag } &=&\cosh \left( r\right) a^{\dag }-\sinh \left( r\right) \exp
\left( -i\theta \right) a  \notag \\
&=&S^{\dag }a^{\dag }S,  \label{relat}
\end{eqnarray}%
where the squeeze operator $S=\exp [\frac{1}{2}(\xi ^{\ast }a^{2}-\xi
a^{\dagger 2})]$. Using this squeezed mode we can rewrite Eq. (\ref{master})
as%
\begin{equation}
\dot{\rho}=\frac{\gamma (t)}{2}(2A\rho A^{\dagger }-A^{\dagger }A\rho -\rho
A^{\dagger }A).  \label{trmaster}
\end{equation}%
Substituting Eqs. (\ref{relat}) to Eq. (\ref{trmaster}) and defining $\rho
_{s}=S\rho S^{\dagger \text{ }}$we can obtain%
\begin{equation}
\dot{\rho}_{s}=\frac{\gamma (t)}{2}(2a\rho _{s}a^{\dagger }-a^{\dagger
}a\rho _{s}-\rho _{s}a^{\dagger }a).  \label{mast}
\end{equation}%
$\{a,a^{\dagger },a^{\dagger }a,1\}$ constitutes the algebra$\ hw\left(
4\right) $. Based on the left and right representations of certain algebra
\cite{Wang01}, we define the left and right representations of this algebra
as%
\begin{eqnarray}
hw\left( 4\right) _{r} &=&\left\{ a^{r},a^{r\dagger },n^{r}=a^{r\dagger
}a^{r},1\right\} ,  \notag \\
hw\left( 4\right) _{l} &=&\left\{ a^{l},a^{l\dagger },n^{l}=a^{l\dagger
}a^{l},1\right\} ,  \label{comm1}
\end{eqnarray}%
where $hw\left( 4\right) _{r}$ acts to the right on the ket state $|n\rangle
$ and $hw\left( 4\right) _{l}$ acts to the left on the bra state $\langle n|$%
. They have the commutation rules
\begin{eqnarray*}
\lbrack a^{r},a^{r\dagger }] &=&1,[a^{r},n^{r}]=a^{r},[a^{r\dagger
},n^{r}]=-a^{r\dagger }; \\
\lbrack a^{l},a^{l\dagger }] &=&-1,[a^{l},n^{l}]=-a^{l},[a^{l\dagger
},n^{l}]=a^{l\dagger }.
\end{eqnarray*}%
It is evident that $hw\left( 4\right) _{r}$ ($hw\left( 4\right) _{l}$) is
isomorphic (anti-isomorphic, respectively) to $hw\left( 4\right) $. Since
they act on different spaces, the operators commute with each other%
\begin{equation*}
\lbrack hw\left( 4\right) _{r},hw\left( 4\right) _{l}]=0.
\end{equation*}

Having these useful algebras at hand, we can constitute the composite
algebra $C$,%
\begin{equation*}
C=\{K_{+}=a^{r\dagger }a^{l},K_{-}=a^{r}a^{l\dagger },K_{0}=\frac{n^{r}+n^{l}%
}{2}\}.
\end{equation*}%
We see that $C$ is an $su(1,1)$ algebra from the commutation rules%
\begin{equation*}
\lbrack K_{0},K_{\pm }]=\pm K_{\pm },[K_{-},K_{+}]=2K_{0}
\end{equation*}%
which can be derived from Eq. (\ref{comm1}). The operators have the
following actions on the base of von Neumann space as%
\begin{eqnarray}
K_{+}|n\rangle \langle m| &=&\sqrt{\left( n+1\right) \left( m+1\right) }%
|n+1\rangle \langle m+1|,  \notag \\
K_{-}|n\rangle \langle m| &=&\sqrt{nm}|n-1\rangle \langle m-1|,  \notag \\
K_{0}|n\rangle \langle m| &=&\frac{n+m+1}{2}|n\rangle \langle m|.
\label{act}
\end{eqnarray}%
Then the master equation (\ref{master}) can be converted into a Schr\"{o}%
dinger-like equation%
\begin{eqnarray}
\dot{\rho}_{s} &=&\Gamma \left( t\right) \rho _{s},  \notag \\
\Gamma \left( t\right) &=&\gamma (t)(K_{-}-K_{0}+\frac{1}{2}).
\label{master3}
\end{eqnarray}%
So the quantum master equation (\ref{master}) possesses the $su\left(
1,1\right) $ dynamical symmetry. It is thus integrable and can be solved
analytically according to the algebraic dynamics \cite{Wang93}.

\bigskip

To study the property of steady states, we should solve the eigen equation
of the master equation for the steady case $\gamma (t)\rightarrow const.$%
\begin{equation}
\Gamma \rho _{s}=\beta \rho _{s}.  \label{eigen}
\end{equation}%
Introducing a similarity transformation%
\begin{equation*}
U=e^{-K_{-}}
\end{equation*}%
we can diagonalize the operator $\Gamma $ as%
\begin{equation}
\bar{\Gamma}=U^{-1}\Gamma U=\gamma (-K_{0}+\frac{1}{2}).  \label{trans}
\end{equation}%
From Eq. (\ref{trans}), we get the solution of Eq. (\ref{eigen}) as follows%
\begin{eqnarray*}
\beta _{nm} &=&\frac{\gamma }{2}\left( n+m\right) ; \\
\rho _{s} &=&e^{-K_{-}}|n\rangle \langle m|.
\end{eqnarray*}%
Then unique steady solution of the master equation (\ref{master}), $\rho
_{s0}=e^{-K_{-}}|0\rangle \langle 0|=|0\rangle \langle 0|$ which is the
zero-mode solution $(\beta _{00}=0)$ of the eigen equation (\ref{eigen}),
corresponds to the squeezed state of the field:
\begin{equation}
\rho _{0}=S^{\dagger }\rho _{s0}S=S^{\dagger }|0\rangle \langle 0|S.
\label{stead}
\end{equation}

Next, we study the time-dependent solution of the nonautonomous
master equation (\ref{trmaster}). To this end, we go back to Eq. (\ref%
{master3}) and introduce the time-dependent gauge transformation%
\begin{equation*}
U_{g}\left( t\right) =e^{\alpha \left( t\right) K_{-}}.
\end{equation*}%
Under the gauge transformation condition
\begin{equation}
\dot{\alpha}\left( t\right) =\gamma (t)[\alpha \left( t\right) +1],
\label{gtr}
\end{equation}%
the rate operator $\Gamma \left( t\right) $ can be diagonalized as%
\begin{equation*}
\bar{\Gamma}(t)=U_{g}^{-1}\left( t\right) \Gamma U_{g}\left( t\right)
-U_{g}^{-1}\left( t\right) \dot{U}_{g}\left( t\right) =\gamma \left(
t\right) (-K_{0}+\frac{1}{2}).
\end{equation*}%
Then the transformed equation reads as%
\begin{eqnarray*}
\frac{d\bar{\rho}_{s}(t)}{dt} &=&\bar{\Gamma}(t)\bar{\rho}_{s}(t), \\
\bar{\rho}_{s}(t) &=&U_{g}^{-1}\left( t\right) \rho _{s}(t).
\end{eqnarray*}%
So the solution of Eqs. (\ref{master3}) can be obtained readily as%
\begin{eqnarray}
\rho _{s}\left( t\right) &=&U_{g}\left( t\right) e^{\int_{0}^{t}\bar{\Gamma}%
(\tau )d\tau }\rho _{s}(0),  \notag \\
&=&\sum_{m,n}C_{m,n}e^{-\frac{n+m}{2}\int_{0}^{t}\gamma \left( \tau \right)
d\tau }e^{\alpha \left( t\right) K_{-}}|n\rangle \langle m|,  \label{tsolu}
\end{eqnarray}%
where we have used the gauge condition $U_{g}\left( 0\right) =1$ and the
initial state $\rho _{s}(0)=\sum_{m,n}C_{m,n}|n\rangle \langle
m|(\sum_{m}C_{m,m}=1)$. After the inverse squeezing transformation we obtain
the solution of Eq. (\ref{master})%
\begin{equation}
\rho \left( t\right) =\sum_{m,n}C_{m,n}e^{-\frac{n+m}{2}\int_{0}^{t}\gamma
\left( \tau \right) d\tau }S^{\dagger }\{e^{\alpha \left( t\right)
K_{-}}|n\rangle \langle m|\}S=\sum_{m,n}C_{m,n}\rho _{nm}\left( t\right) .
\label{solu}
\end{equation}

In the following we shall prove that the time-dependent solution (\ref{solu}%
) will approach the steady solution (\ref{stead}) asymptotically. To study
the asymptotical behavior of $\alpha \left( t\right) $, we define $%
y(t)=\alpha \left( t\right) \exp [-\int_{0}^{t}\gamma \left( \tau \right)
d\tau ]$, the time differential of $y(t)$ is given by $b\left( t\right) \exp
[-\int_{0}^{t}\gamma \left( \tau \right) d\tau ]$ where $b(t)=\dot{\alpha}%
\left( t\right) -\gamma (t)\alpha \left( t\right) =\gamma (t)$. Because $%
b\left( t\right) \rightarrow $ $\gamma \left( \infty \right) =\gamma $ is
bounded for large $t$, the differential $\overset{\cdot }{y}\left( t\right) $
tends to zero. Hence $y(t)$ is towards to a constant. This implies that $%
\alpha \left( t\right) $ diverges asymptotically. Therefore one has
the asymptotical properties of Eq. (\ref{gtr})
\begin{eqnarray}
y(t)|_{t\rightarrow \infty } &=&const,  \notag \\
\alpha \left( t\right) |_{t\rightarrow \infty } &=&\infty .  \label{asy}
\end{eqnarray}%
Using Eq. (\ref{act}) we can rewrite any component of Eq. (\ref{solu}) as
\begin{equation*}
\rho _{nm}\left( t\right) =S^{\dagger }\{y(t)^{\frac{^{n+m}}{2}}\sum_{q=|%
\frac{n-m}{2}|}^{\frac{n+m}{2}}\alpha (t)^{-q}\frac{\sqrt{n(n-1)\cdots (q+%
\frac{n-m}{2}+1)m(m-1)\cdots (q-\frac{n-m}{2}+1)}}{(\frac{n+m}{2}-q)!}|q+%
\frac{n-m}{2}\rangle \langle q-\frac{n-m}{2}|\}S.
\end{equation*}%
Substituting the asymptotical conditions Eqs. (\ref{asy}) into the
above equation, we can see that only the term corresponding to $n=m$
and $q=0$ will survive, yielding the unique equilibrium steady
state. Therefore, any
time-dependent solution (\ref{solu}) of the nonautonomous master equation (%
\ref{master}) tends to the unique steady solution (\ref{stead})
asymptotically,%
\begin{equation*}
\rho _{nm}\left( t\right) \rightarrow \delta _{nm}\rho _{0}=S^{\dagger
}|0\rangle \langle 0|S\delta _{nm},\rho \left( t\right) \rightarrow \rho
_{0}=S^{\dagger }|0\rangle \langle 0|S.
\end{equation*}

Since any initial state approaches the steady solution which is the squeezed
state of the single mode field, one comes to the important and interesting
conclusion that the squeezed vacuum field reservoir always brings any
initial state of the\ single mode field to its squeezed state
asymptotically, and the squeezed vacuum field reservoir plays the role of a
squeezing mold which transforms any kind of monomode cavity field state into
its squeezed state: the squeezed vacuum imprints its squeezing information
to the cavity field exclusively. This property may be useful in practice to
generate a squeezed state of a cavity field.

\end{document}